\documentclass[reprint, % for submission, 2 columns
 amsmath,amssymb,
 aps,
 prab,
floatfix,
showkeys
]{revtex4-2}
\usepackage{natbib}
\usepackage{dcolumn}% Align table columns on decimal point
\usepackage{bm}
\usepackage[utf8]{inputenc} % allow utf-8 input
\usepackage[T1]{fontenc}    % use 8-bit T1 fonts
\usepackage{hyperref}       % hyperlinks
\usepackage{url}            % simple URL typesetting
\usepackage{booktabs}       % professional-quality tables
\usepackage{amsfonts}       % blackboard math symbols
\usepackage{nicefrac}       % compact symbols for 1/2, etc.
\usepackage{microtype}      % microtypography
\usepackage{graphicx}
\usepackage{amsthm}
\usepackage{mathtools}
\usepackage{siunitx}
\usepackage{tikz}
\usepackage[caption=false]{subfig}
\usetikzlibrary{positioning,chains}

\begin{document}
%\preprint{APS/123-QED}

\title{Review of Time Series Forecasting Methods\texorpdfstring{\\}{}and Their Applications to Particle Accelerators}% Force line breaks with \\

\author{
\href{https://orcid.org/0000-0003-4881-2166}{\includegraphics[scale=0.06]{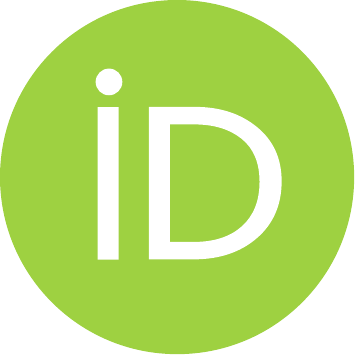}\hspace{1mm}Sichen~Li}}%
\altaffiliation[Also at the ]{Department of Physics, ETH Zurich}
\email{sichen.li@psi.ch}
\author{\href{https://orcid.org/0000-0002-7230-7007}{\includegraphics[scale=0.06]{figs/orcid.pdf}\hspace{1mm}Andreas~Adelmann}}%
\email{andreas.adelmann@psi.ch}
\affiliation{%
Paul Scherrer Institute\\
5232 Villigen\\
Switzerland%\\
}%

\date{\today}% It is always \today, today,

%%% Add PDF metadata to help others organize their library
%%% Once the PDF is generated, you can check the metadata with
%%% $ pdfinfo template.pdf
\hypersetup{
pdftitle={Review of time series forecasting methods and their applications to particle accelerators},
pdfsubject={},
pdfauthor={Sichen~Li, Andreas~Adelmann},
pdfkeywords={Machine learning, particle accelerator, time series forecasting},
}

\begin{abstract}
    Particle accelerators are complex facilities that produce large amounts of structured data and have clear optimization goals as well as precisely defined control requirements. As such they are naturally amenable to data-driven research methodologies. The data from sensors and monitors inside the accelerator form multivariate time series. With fast pre-emptive approaches being highly preferred in accelerator control and diagnostics, the application of data-driven time series forecasting methods is particularly promising.
    
    This review formulates the time series forecasting problem and summarizes existing models with applications in various scientific areas. Several current and future attempts in the field of particle accelerators are introduced. The application of time series forecasting to particle accelerators has shown encouraging results and the promise for broader use, and existing problems such as data consistency and compatibility have started to be addressed.
\end{abstract}

% keywords can be removed
\keywords{Machine learning; particle accelerator; time series forecasting}

\maketitle

\section{Introduction}
% general introduction
Particle accelerators have a significant role in various areas of science, from searches for new physics and nuclear-waste transmutation to cancer treatment. They are also facilities that lend themselves to data-driven research methodologies, such as event forecasting based on machine learning (ML), given that they produce substantial volumes of structured data and that their operation is defined by clear optimization goals and precise control requirements. To achieve optimal operational conditions while keeping the accelerator under control and within safety limits, a multitude of different sensors and monitors are placed at specific positions inside the accelerator complex. The data are recorded as multivariate time series, sampled at specified frequencies. The future values of some quantities of interest, or a potential failure of the machine, might then be inferred by time series forecasting methods.

% problem formulation

A \textit{time series} $\vec{x}_t \in \mathbb{R}^n$, where $t$ stands for time and $n$ is the dimension of the desired variables, is ``a collection of observations made sequentially through time''~\cite{chatfield2000time}. The \textit{forecasting problem} is to infer the future value $\vec{x}_{t+h}$ based on the current and past values of $\vec{x}$, where $h$ is called the \textit{lead time}. Typical examples of $\vec{x}_t$ in a particle accelerator scenario include the measurement of beam current, magnet strength and temperature, and the output of loss monitors.

According to the dimension $n$ of the input signals $\vec{x}$, forecasting problems can be categorized into \textit{univariate} and \textit{multivariate} problems. Based on the value of $h$, they can also be divided into \textit{one-step-ahead} or \textit{multi-step-ahead} problems. Section~\ref{sec:method} introduces existing methods in two main categories: \emph{linear} and \emph{nonlinear} models. The illustration of each method is accompanied by practical applications in fields such as energy and finance. This review paper chooses to focus on several typical, commonly used and prospective methods, especially those of interest for applications in particle accelerator diagnostics. 

In practical terms, the quantity of interest in particle accelerator operation is sometimes not only the future values of the input signals $\vec{x}_{t+h}$, but rather another value depending on $\vec{x}_{t+h}$, i.e., $y_{t+h} = f(\vec{x}_{t+h})$. An example is the probability of machine failure in $h$ seconds following the latest measurements, where $y \in [0,1]$. In this example, the form of the function $f$ needs also to be inferred. Such a problem set-up of learning $f$ fits into the scope of \emph{anomaly detection}, where an anomaly score $y_t$ (usually $y_t \in [0,1]$, or a non-negative value) is inferred from input $\vec{x}_t$ at every timestamp $t$. However, the combined problem --- extrapolating from the input time series $\vec{x}_{t}$ to the future anomaly score $y_{t+h}$ --- is formulated rather ambiguously in current studies~\cite{fahim2019anomaly,lin2020anomaly}. Section~\ref{sec:accel} lists several existing attempts in the particle accelerator field aiming to tackle such composite `anomaly forecasting' problems. In this context, also the topic of remaining useful life (RUL) predictions in predictive maintenance is discussed, which is introduced in Section~\ref{sec:disc}.

\section{Forecasting methods}\label{sec:method}
According to the basic assumptions about the underlying generating process, time series models can be categorized into \emph{linear} and \emph{nonlinear} ones. In the former category, we introduce the auto-regressive integrated moving average (ARIMA) and state-space models. In the latter category, after introducing the general concept of artificial neural network (ANN),  we start from the simple multilayer perceptron (MLP), and then move towards the more complex recurrent neural network (RNN). Recent attempts in hybrid models that integrate the linear and nonlinear models are also introduced and shortly discussed.

\subsection{Linear models}

\subsubsection{ARIMA}

The \emph{auto-regressive integrated moving average} (ARIMA) class of models, introduced by \citet{box2015time}, laid the foundation for many variations and further developments in time series forecasting. It assumes that the prediction is a weighted linear sum of past observations and random errors. A univariate auto-regressive moving average (ARMA) model of order $(p, m)$ follows the relation
\begin{equation*}
    x_t = \theta_0 + \epsilon_t + \sum_{i=1}^{p}\phi_i x_{t-i}+ \sum_{j=1}^{m}\theta_j \epsilon_{t-j}
\end{equation*}
where $\epsilon_t$ is the random error, which is assumed to be independently and identically distributed with mean $\mu=0$ and variance $\sigma^2$; $\phi_i, i=0 \dots p$ and $\theta_j, j=0 \dots m$ are model parameters, with $\theta_0$ indicating a constant contribution~\cite{zhang2003time}. It combines the previously formulated concept of auto-regressive (AR) process from~\citet{udny1927method} and~\citet{walker1931periodicity}, and moving average (MA) techniques from pioneering works of~\citet{allen1950work}. Explicitly, by setting $m=0$,
\begin{equation*}
    x_t = \theta_0 + \epsilon_t + \sum_{i=1}^{p}\phi_i x_{t-i}
\end{equation*}
becomes an AR model of order $p$. And setting $p=0$,
\begin{equation*}
    x_t = \theta_0 + \epsilon_t +\sum_{j=1}^{m}\theta_j \epsilon_{t-j}
\end{equation*}
returns a MA model of order $m$~\cite{chatfield2000time}.
Then the ``I'' in ARIMA means ``integrated'', which refers to the operation of differencing, which converts an originally non-stationary time series into a stationary one. The general form of an ARIMA model is denoted ARIMA$(p,d,m)$, where $d$ refers to the number of differencing steps needed to reach stationarity.

% Exponential smoothing
Other approaches such as general exponential smoothing (GES), which was originally introduced by~\citet{brown1961fundamental} and~\citet{holt1960planning} and uses multiples of polynomials, sinusoids and exponentials of time to model the trend, could also be incorporated in ARIMA. Furthermore, according to~\citet{gardner1985exponential}, ``the equivalent ARIMA model is even simpler and more efficient'' than GES in standard form. The multi-dimensional generalization of ARIMA extends the univariate $x_t$ into a set of $n$ interrelated variables $\vec{x}_t=(x_{1,t}, \dots, x_{n,t})$, leading to the vector-ARIMA (VARIMA) method, where each component of $\vec{x}$ is modelled as a linear sum of present and past values of all $n$ components and a multivariate white noise. Such a problem is often referred to as \emph{multiple time-series modelling}~\cite{chatfield2000time}.
%, where the $n$ variables are called as ``arising on an equal footing''

In addition to the model formulation, Box and Jenkins have also established a practical approach to build ARIMA models --- now known as the \emph{Box--Jenkins approach} --- that strings together model identification, estimation, and verification into a full iterative cycle~\cite{hipel1977advances}. After removing potential non-stationarity and seasonality through differencing~\cite{dickey1987determining}, a plausible model of the orders $p$ and $m$ is identified from checking autocorrelation patterns or other model-selection criteria of the time series. Then the model parameters $\theta$ and $\phi$ are fitted by minimising the overall errors. Finally, various diagnostic checks are performed on the residual of the real series and the fitted model. The three-step cycle is typically run several times before reaching a satisfactory model. The versatility of ARIMA usually enables it to imitate time series of diverse types, without having to introduce many parameters.%~\cite{de200625}.

Ever since its proposal, the ARIMA model has had a key role in a wide range of forecasting-related areas, including the recent application in predictions of the Covid-19 epidemic evolution~\cite{benvenuto2020application}. \citet{de200625} provide a comprehensive list of earlier empirical applications of ARIMA and its variants in the scope of \textit{International Journal of Forecasting} papers. More recently, the ARIMA model is serving increasingly as one of the comparison baselines for newly developed nonlinear models. As an example, \citet{siami2018comparison} show the superiority of the long short-term memory (LSTM) model over ARIMA on several standard time series datasets. However, such superior performance does not challenge ARIMA's position as the foundation of forecasting models. Considering its robustness, high interpretability as well as the black-box nature of many models based on deep learning~\cite{makridakis2018statistical}, ARIMA appears increasingly as a fundamental component in hybrid models, which take advantage of its statistical properties while avoiding its linear rigidity. \citet{zhang2003time} proposes to combine the forecasts from a linear ARIMA model and a nonlinear artificial neural network (ANN) model. On this basis, \citet{wang2013arima} test the combination of ARIMA and ANN in both additive and multiplicative ways, and the latter shows consistent improvement of forecasting accuracy compared to ARIMA and ANN individually as well as to the additive hybrid model. One notable application is that by~\citet{liu2021application} on wind speed prediction, where they employ an empirical mode-decomposition approach that uses ARIMA for low-frequency and LSTM for high-frequency sub-sequences prediction.

\subsubsection{State-space model}

A second important and practical formulation is based on \textit{linear dynamical systems}. It develops a recursive algorithm for computing forecasts. Originating from control engineering, the model views any observation at time $t$ as a signal part plus a noise part, and the signal is then decomposed into a linear combination of $q$ \emph{state variables}, to form the state vector $\vec{h}_t \in \mathbb{R}^q$. The \emph{observation} (or \emph{measurement}) equation for univariate $x_t$ reads
\begin{equation}
    x_t = f(\vec{h}_t) + \epsilon_t \label{eq:state1}
\end{equation}
where $f$ is a function and $\epsilon_t$ denotes the zero-mean noise part of time series $x_t$. The future values of the state vector $\vec{h}_t$ only depends on its current value, and not on its past --- in other words, the state vector has \emph{Markovian} properties. In linear state-space models, $\vec{h}_t$ is assumed to evolve according to the \emph{transition} equation
\begin{equation}
    \vec{h}_t = \mathbf{G}_t \vec{h}_{t-1} + \vec{\xi}_t \label{eq:state2}
\end{equation}
where $\mathbf{G}_t$ is a $q\times q$ transition matrix, and $\vec{\xi}_t \in \mathbb{R}^q$ is the disturbance term of the state vector $\vec{h}_t$, assumed to have zero mean. State-space models make fewer assumptions about the form of the trend, yet they still can produce adaptive and robust forecasts~\cite{chatfield2000time}. For instance, \citet{bae1995comparison} show better performance of the state-space model in both cycle tracking and error reduction relative to plain multiple regression in short-term multivariate forecasts of U.S. fuel consumption.

The corresponding updating procedure of the state-space model is the so-called \textit{Kalman filter}~\cite{kalmanoriginal}, which recursively updates the estimate of the state vector $\vec{h}_t$ and thereby calculates the latest forecast $x_t$ whenever a new observation becomes available. For instance, in the case of a one-dimensional state space with $f$ the identity function and constant $G_t=G$, we have
\begin{align}
    x_t &= h_t + \epsilon_t \nonumber\\
    h_t &= G \cdot h_{t-1} + \xi_t \label{eq:state}
\end{align}
where we assume $\epsilon_t \sim \mathcal{N}(0, \sigma^2)$ and $\xi_t \sim \mathcal{N}(0, \tau^2)$. \citet{kalmancourse} gives an introductory example of the Kalman-filter algorithm with detailed derivation. It starts from an initial guess of the mean and variance of the initial state $h_0$, denoted by $h_0^0$ and $P_0^0$. The upper index refers to the number of observations that have been used to update the state, and the lower index refers to the time $t$. Following Eq.~\eqref{eq:state}, we can calculate our guess for the mean and variance of the next state $h_1$
\begin{align*}
    h_1^0 &= G \cdot h_0^0 \quad\text{(mean)}\\
    P_1^0 &= G^2 \cdot P_0^0 + \tau^2 \quad\text{(variance)}
\end{align*}
With the observation $x_1$ as new information, we can update the previous guess $h_1^0$ and $P_1^0$ to $h_1^1$ and $P_1^1$:
\begin{align*}
    h_1^1 &= h_1^0 + K_1\cdot(x_1 - h_1^0) \\
    P_1^1 &= (1-K_1)\cdot P_1^0
\end{align*}
where $K_1 = \frac{P_1^0}{P_1^0 + \sigma^2}$ is the \emph{Kalman gain coefficient} at $t=1$. In general, given the current estimate $h_t^t$ and $P_t^t$ at time $t$ after updating with $t$ observations, we can update with the new observation $x_{t+1}$ according to
\begin{align*}
    h_{t+1}^t &= G \cdot h_t^t \\
    P_{t+1}^t &= G^2 \cdot P_t^t + \tau^2 \\
    K_{t+1} &= \frac{P_{t+1}^t}{P_{t+1}^t + \sigma^2} \\
    h_{t+1}^{t+1} &= h_{t+1}^t + K_{t+1}\cdot(x_{t+1} - h_{t+1}^t) \\
    P_{t+1}^{t+1} &= (1-K_
    {t+1})\cdot P_{t+1}^t.
\end{align*}
A general ARIMA model can also be recasted in the state-space formulation to apply Kalman filtering and to ease the estimation procedure~\cite{harvey1990forecasting}. With details given by~\citet{meinhold1983understanding}, Kalman filtering ensures that the minimum mean squared estimator (MMSE) of the state vector is obtained in case of normal noise. The flexibility of the procedure has been established in various forecasting problems. \citet{harvey1987applications} has written a chapter with a detailed theoretical presentation of the method, supplemented by various applications in econometrics. \citet{visser1995trend} have proposed a trend regression model that incorporates both deterministic and stochastic trends in a general ARIMA format for climatological data. They then transcribe the model into the state-space format and use Kalman filtering to estimate and evaluate the model parameters. The hybrid approach with neural networks is also highly fruitful here. For instance, \citet{peel2008data} builds a Kalman filter on top of an ensemble of neural network models, where he exploits its advantage of handling multiple input sources simultaneously and its intrinsic ability to filter predictions over time.

% Bayesian
It is crucial to quantify the forecast uncertainty, both for unwrapping the underlying models and for better guiding further operations. Taking this insight into account led to the incorporation of Bayesian inference in forecasting models. One inspiring attempt to extend the capacity of linear models is that of ~\citet{ghosh2007bayesian} on traffic flow forecast, which uses Bayesian inference to fit the model parameters in place of the typical point estimation by residue minimization. 

\subsection{Nonlinear models}
Although the linear models discussed above have the advantages of simple implementation, straightforward interpretability and also good performances in some proven cases~\cite{han2012application}, the true model underlying time series may in reality be nonlinear and therefore difficult to unwrap. Earlier attempts of adapting the linear models to nonlinear behaviour include the bilinear model~\cite{rao1981theory}, which contains nonlinear cross terms of past values and white noise, but is based on structural theories analogous to linear models; and the threshold autoregressive (TAR) model~\cite{tong2012threshold}, which combines piece-wise linear models systematically and reaches global nonlinearity with local linearity. These models are mostly confined with only some specific patterns of nonlinearity, however, which in turn results in their weak performance for general problems~\cite{de1992some}.

% general ML
While the methods introduced above, from linear models like ARIMA to nonlinear attempts like TAR, are rather generally considered as \emph{model-driven}, the following broad category of ML methods are more believed to be \emph{data-driven}, in the sense that they do not necessarily require an explicit form of an underlying model. The term `machine learning' has broad and rather ambiguous meanings, with techniques ranging from ordinary least square methods to deep neural networks with millions of parameters (for instance, the well-known image recognition network \emph{ResNet-50} has more than 23 million trainable parameters~\cite{he2016deep}). According to~\citet{bandara2020forecasting}, traditional \emph{model-driven} methods work better when the data volume is minimal. But nowadays, complex ML models that used to be outperformed by traditional statistical models in forecasting of simple short time series~\cite{crone2011advances,  competitions}, have gradually become dominant in the era of ever-increasing data quantity and quality. The revolutionary changes brought by big-data technology enables us to manage longer and uninterrupted time series. In addition, the access to many interrelated series has opened up novel learning possibilities. While traditional models require us to explicitly write out a specific form of correlations, an ML-based model can naturally come up with combined features from various inputs and exploit cross-series information~\cite{hewamalage2021recurrent}. Below we introduce the basic concept of artificial neural network (ANN) and in particular of recurrent neural network (RNN) architectures, together with some extensions and example applications. 

% ANN
\subsubsection{Multi-layer perceptrons}
An ANN is composed of connected units called neurons or nodes, where linear or nonlinear calculations are performed with numbers transmitted and received through the connections. The concept is to mimic how the neural system of humans works, even though this being a much over-simplified analogy. As the simplest form of ANN, multi-layer perceptrons (MLP), also known as feed-forward neural networks, contain only forward connections between nodes, without loops as in RNNs.  Here we describe an MLP with $N$ hidden layers as
\begin{align*}
    \vec{x}^i &= (x_{t-1}, \dots,x_{t-p}) \in \mathbb{R}^p\\
    \vec{h}^1 &= G^1(\vec{b}^1 + \mathbf{W}^{1, i} \cdot \vec{x}^i), 
    \quad \vec{h}^1 = (h^1_1, \dots, h^1_{q_1})\in \mathbb{R}^{q_1} \\
    \quad &\vdots \\
    \vec{h}^N &= G^N(\vec{b}^N + \mathbf{W}^{N,N-1} \cdot \vec{h}^{N-1}), \quad \vec{h}^N% = (h^N_1, \dots, h^N_{q_N})
    \in \mathbb{R}^{q_N} \\
    x_t &= x^o = G^o(b_o + \vec{w}^{o,N}\cdot\vec{h}^N)
\end{align*}
where \emph{i} and \emph{o} denote \emph{input} and \emph{output}, respectively. The input layer $\vec{x}^i$ has $p$ nodes that take $p$ past values of the series. The $j^{th}$ hidden layer $\vec{h}^j$ has $q_j$ nodes, and the output layer $x^o$ has only one output, which is the current value $x_t$. $\mathbf{W}^{1, i}$ is a $(q_1, p)$ matrix of connection weights from the input layer (of $p$ nodes) to the first hidden layer (of $q_1$ nodes), $\mathbf{W}^{N, N-1}$ are weights from the ${(N-1)}^{th}$ to the $N^{th}$ hidden layer, and $\vec{w}^{o,N} \in \mathbb{R}^{q_N}$ are weights from the $N^{th}$ hidden layer to the output. The vectors $\vec{b}$ are the bias of each layer, and the functions $G$ are the activation functions applied on each layer, with typical choices being sigmoid, hyperbolic tangent or rectified linear unit (ReLU)~\cite{nair2010rectified} functions. Figure~\ref{fig:mlp} visualizes the structure of a MLP.

\tikzset{%
%   every neuron/.style={
%     circle,
%     draw,
%     minimum size=0.1cm
%   },
  every neuron/.style={circle,draw,thick,align=center},
  neuron missing/.style={
    draw=none, 
    scale=2,
    text height=0.3cm,
    execute at begin node=\color{black}$\vdots$
  },
}
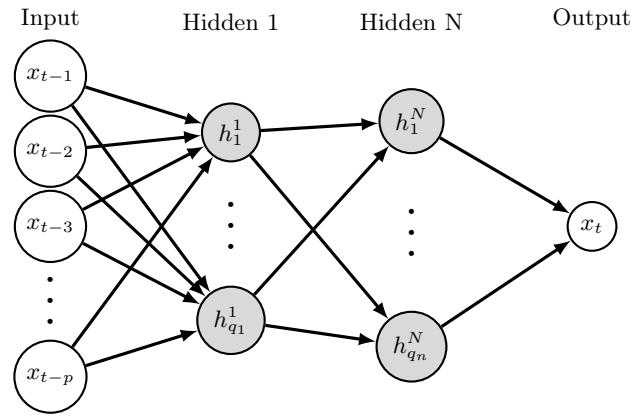
\begin{figure}[b]
\centering
\begin{tikzpicture}[
x=1.2cm, y=1cm] 
%>=stealth]

\foreach \m [count=\y] in {1,2,3,missing,p}
{
    \ifnum\y=4
        \node [every neuron/.try, neuron \m/.try] (input-\m) at (0,2.5-\y) {};
    \else
        \node [every neuron/.try, neuron \m/.try] (input-\m) at (0,2.5-\y) {$x_{t-\m}$};
    \fi
}

\foreach \m/\txt/\cf [count=\y] in {1/$h^1_1$/gray!30,missing//none,2/$h^1_{q_1}$/gray!30}
\node [every neuron/.try, neuron \m/.try,fill=\cf] (hidden1-\m) at (2,2-\y*1.25) {\txt};

\foreach \m/\txt/\cf [count=\y] in {1/$h^N_1$/gray!30,missing//none,2/$h^N_{q_n}$/gray!30}
\node [every neuron/.try, neuron \m/.try,fill=\cf] (hidden2-\m) at (4,2.4-\y*1.5) {\txt};

\node [every neuron/.try, neuron 1/.try] (output-1) at (6,-0.5) {$x_t$};

\foreach \i in {1,2,3,p}
  \foreach \j in {1,...,2}
    \draw [very thick, -latex] (input-\i) -- (hidden1-\j);

\foreach \i in {1,...,2}
  \foreach \j in {1,...,2}
    \draw [very thick, -latex] (hidden1-\i) -- (hidden2-\j);
    
\foreach \i in {1,...,2}
\draw [very thick, -latex] (hidden2-\i) -- (output-1);

\foreach \l [count=\x from 0] in {Input, Hidden 1, Hidden N, Output}
  \node [align=center, above] at (\x*2,2) {\l};

\end{tikzpicture}
\caption{Illustration of a MLP.}
\label{fig:mlp}
\end{figure}

The advantages of ANNs go beyond the aforementioned lenient requirement for a priori assumptions; they can universally approximate various forms of functions~\cite{hornik1989multilayer}, and perform especially well when modelling nonlinear relations. They also possess high generalization power towards out-of-sample data.

\citet{zhang1998forecasting} give a detailed overview of ANNs for forecasting problems, with the focus on MLP and some hybrid approaches. Starting from the earliest success by~\citet{lapedes1987nonlinear} on a generated dataset following deterministic nonlinear dynamics, ANNs have been widely applied to various fields. In finance, the inherent noise, fat-tail distributions and nonlinear patterns of the financial data make it difficult for conventional methods such as ARIMA to capture the dynamics, which lead to the popularity of ANNs. \citet{li2010applications} comprehensively introduce the application of ANNs on problems ranging from microscopic ones such as exchange-rate or stock-price forecasting to macroscopic scenarios such as financial-crisis forecasting, while~\citet{krollner2010financial} made a more specialised survey on ANN-related applications in stock-index forecasting. 

\subsubsection{Recurrent neural networks}
One drawback of the MLP approach is that it isolates the inputs at every timestamp and treats them as independent variables. The temporal order and the dependencies of the input series on time contain crucial information about the evolution of the series, but are not taken into account~\cite{bianchi2017overview}. This issue is targeted and overcome by RNNs, which by design can carry forward --- or in a more neurological term, `remember' --- the states from previous inputs.

The unit component that constitutes a RNN is called an \emph{RNN cell}. A generic RNN cell consists of the \emph{input} time series $x_t$ (in the univariate case), a \emph{hidden state} $\vec{h}_t \in \mathbb{R}^q$ with $q$ the cell dimension, and the \emph{cell output} $\vec{o}_t \in \mathbb{R}^q$. In the case of time series forecasting, the cell output $\vec{o}_t$ needs to be transformed again into the final network output $x_{t+h}$, which is the $h$-step-ahead prediction of $x_t$, normally by fully connected layers. At each time step, the hidden state $\vec{h}_t$ is updated according to Eq.~\eqref{eq:rnn1}, where $G$ is a nonlinear activation function, $\vec{o}_t$ is computed according to $\vec{h}_t$ and activation function $G^o$, and $x_{t+h}$ is the final prediction. 

\begin{align}\label{eq:rnn1}
\quad&\left.
\begin{aligned}
\vec{h}_t &= G(\vec{h}_{t-1}, \vec{x}_t, W^i, V^i, b^i) \nonumber\\
\vec{o}_t &= G^o(\vec{h}_t, W^o, b^o)
\end{aligned} \right\} \text{inside RNN cell} \\
&x_{t+h} = f(\vec{o_t}) \quad\quad \text{outside RNN cell}
\end{align}

In Eq.~\eqref{eq:rnn1}, $W^i, W^o$ are weights associated with the hidden state, $V^i$ are weights of the input, and $b^i, b^o$ are bias. Here \emph{i} and \emph{o} again denote \emph{input} and \emph{output}. Their similar forms with Eq.~\eqref{eq:state1} and \eqref{eq:state2} reveal that RNNs could be interpreted as a kind of nonlinear state-space model from a time-series perspective~\cite{masini2021machine}.

\begin{figure}[b]
\centering
\begin{tikzpicture}[item/.style={circle,draw,thick,align=center,scale=0.7},
% cell/.style={% For the main box
%     rectangle, 
%     rounded corners=5mm,
%     draw,
%     %very thick,
%     dashed
%         },
itemc/.style={item,on chain,join}]
 \begin{scope}[start chain=going right,nodes=itemc,every
 join/.style={-latex,very thick},local bounding box=chain]
 \path node[fill=gray!30] (A0) {$\vec{h}_0$} node[fill=gray!30] (A1) {$\vec{h}_1$} node[fill=gray!30] (A2) {$\vec{h}_2$} node[xshift=2em,fill=gray!30] (At)
 {$\vec{h}_t$};
 \end{scope}
 \node[left=1em of chain,scale=2] (eq) {$=$};
 \node[left=2em of eq,item,fill=gray!30] (AL) {$\vec{h}_t$};
 \path (AL.west) ++ (-1em,2em) coordinate (aux);
 \draw[very thick,-latex,rounded corners] (AL.east) -| ++ (1em,2em) node [below, midway] (bi) {$b^i$} -- (aux) 
 |- (AL.west) node [below, midway] (Wi) {$W^i$};
 \node[above=2em of Wi]{$W^o$};
 \node[above=2em of bi]{$b^o$};
 \foreach \X in {0,1,2,t} 
 {\draw[very thick,-latex] (A\X.north) -- ++ (0,2em)
 node[above,item,fill=gray!30] (o\X) {$\vec{o}_\X$};
 \draw[very thick,latex-] (A\X.south) -- ++ (0,-2em)
 node[below, item] (x\X) {$x_\X$};}
\draw[very thick,-latex] (o0.north) -- ++ (0,2em)
 node[above] (y0) {$x_h$};
\foreach \X in {1,2,t} 
{\draw[very thick,-latex] (o\X.north) -- ++ (0,2em)
node[above] (y\X) {$x_{\X+h}$};
}
 \draw[white,line width=0.8ex] (AL.north) -- ++ (0,1.9em);
 \draw[very thick,-latex] (AL.north) -- ++ (0,2em)  %node[below left]{$W^o$}
 %node[below right]{$b^o$}
 node[above,item,fill=gray!30] (oL) {$\vec{o}_t$};
 \draw[very thick,-latex] (oL.north) -- ++ (0,2em) node[right, midway]{$f$}
 node[above] (yL) {$x_{t+h}$};
 \draw[very thick,latex-] (AL.south) -- ++ (0,-2em)
 node[below,item] {$x_t$} node [right, midway] {$V^i$};
 \path (x2) -- (xt) node[midway,scale=2,font=\bfseries] {\dots};
% \node [cell, minimum height =4cm, minimum width=6cm] at (0,0){} ;
\end{tikzpicture}
\caption{Illustration of a RNN cell in folded and unfolded version.}
\label{fig:rnn}
\end{figure}
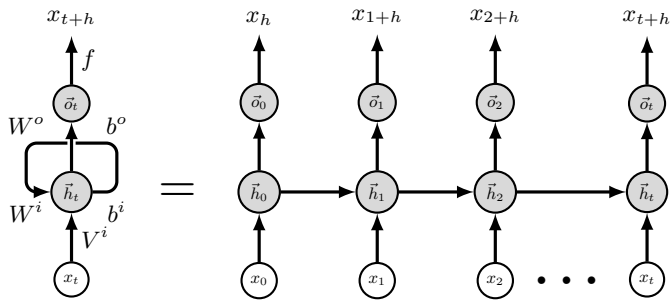

Figure~\ref{fig:rnn} shows the propagating of a generic RNN cell in looped and unfolded view. It demonstrates the feedback structure that enables the network to propagate past states into future time steps, and highlights how the learning process naturally follows the evolution of the series.

Three types of RNN cell are most widely used, especially in forecasting: the basic Elman RNN (ERNN) cell~\cite{elman1990finding}, the gated recurrent unit (GRU) cell~\cite{cho2014learning}, and the LSTM cell~\cite{hochreiter1997long}. Each type of cell has its own formulation and updates the hidden state differently. For instance, an example ERNN cell with the sigmoid function $\sigma$ and the hyperbolic tangent function $\tanh$ as activation functions has the following updating scheme
\begin{align*}
    \vec{h}_t &= \sigma(W^i\vec{h}_{t-1}+V^i\vec{x}_t+b^i), \\
    \vec{o}_t &= \tanh{(W^o\vec{h}_t + b^o)}.
\end{align*}

% seasonality
According to~\citet{hewamalage2021recurrent}, RNNs are suitable for modelling time series with homogeneous seasonal patterns; otherwise, the seasonality of the series needs to be properly handled. \citet{suilin2017kaggle} addressed this by using attention weights, with the idea that if a series possesses daily periodicity, it is beneficial to assign more weight to the value one day ago. This might be a useful technique in particle accelerator scenario where periodicity is either observed or even required. 

% new trend
New developments have brought an innovative trend that further integrates the deep neural network methodology --- which possesses strong learning capability --- with the traditional models, which are more stable and interpretable. Compared to the above-mentioned hybrid models where the final output is simply composed of outputs from multiple separate sub-models, recent novel architectures aim to merge from a more fundamental level. The DeepAR algorithm is capable of producing accurate probabilistic forecasts by training an auto-regressive recurrent neural network on multivariate related time series~\cite{salinas2020deepar}. Furthermore, the deep state-space model~\cite{rangapuram2018deep} parametrizes a linear state-space model per time series with a jointly learned recurrent neural network. As a consequence, desired properties from both sides are satisfied: data efficiency and interpretability from state space models, and the ability to learn complex patterns from deep learning approaches.

\section{Application in particle accelerator diagnostics}\label{sec:accel}
Recent years have seen a boost in applications of data-driven methods in theoretical and engineering research around particle accelerators. Among all promising areas of application, such as beam dynamics modelling~\cite{zhao2020beam} and beam energy optimization~\cite{kirschner2019bayesian, swissfelprab}, diagnostics and control has always been an indispensable and crucial part in ensuring normal operation of the accelerator. A recent application of the Kalman filter technique in a particle accelerator use case was presented by~\citet{syed2021koopman} from the European XFEL, where they applied it for anomaly detection of superconducting RF cavities. By introducing the Koopman operator, they achieved a speed-up of three orders of magnitudes with a linear approximation of the previous nonlinear state-space model. However, attempts in this area that frame such applications as forecasting problems are not yet emerging widely. Due to the complex nature of the data and large diversities across different accelerators, the problem of predicting the future behaviour of an accelerator facility is not as straightforward to formulate as that of a degrading engine.

Here we first present four studies from the Spallation Neutron Source (SNS) at Oakridge, the High Intensity Proton Accelerators (HIPA) at the Paul Scherrer Insitute, the Advanced Photon Source (APS) at Argonne National Laboratory and the Continuous Electron Beam Accelerator Facility
(CEBAF) at Jefferson Lab. They all focus on failure prediction from an anomaly detection perspective, explore existing models or attempt to establish new models, thereby opening up possibilities for future research. The goal of all four studies is to identify potential anomalous events in advance, which embeds a different concept of \emph{forecasting} than the previously introduced models. Then we introduce a newly published study from CERN that implements \emph{autoregressive} modelling for beam loss prediction to cope with machine drift on the timescale of years. Following a similar direction, a diagnostic and feedback system that utilizes time evolution to reduce energy deviation is also planned at the HiRES beamline at the Lawrence Berkeley National Laboratory.

% anomaly detection attempts

\subsection{Pre-emptive detection of machine trips at the Spallation Neutron Source}

The accelerator system of SNS delivers proton pulses of \SI{}{\micro\second} timescale to a steel target, for the production of neutrons through the spallation process. Each of the beam loss trips in SNS costs around \SI{40}{\second} down time, which amounts to about 33,000 lost pulses daily. If such failure could be predicted in advance, the machine protection system (MPS) could react to it by suspending the beam production and resetting the machine, which would in turn reduce the down time to \SI{1}{\second} each. In addition, the reduction of beam loss trips could also lower the damage to the superconducting cavities and reduce the radioactivation of the accelerator. Compared to existing methods that identify the machine trips, \citet{revsvcivc2022improvements} aim to provide an approach that is not only generalizable and system-agnostic across all subsystems and machines, but also pre-emptive, in that it should predict failures in advance instead of reporting them after they have already occurred.

The data used in this work are univariate pulses taken from the SNS differential beam current monitor (DCM) in March 2021. Each pulse is a waveform of 120,000 data points at a frequency of \SI{100}{\mega\hertz}. The length between two consequent pulses is about \SI{16}{\milli\second} during normal operation, which is referred to as the \emph{time budget} allowed to make predictions. The problem is formulated as binary classification, where the pulses before the machine trips are labeled as \emph{bad} pulses. When the model outputs a \emph{bad} label that indicates a potential failure, the actual trip would then come after the current pulse, and in this way forecasting is realized.

\begin{figure*}
    \centering
    \includegraphics[width=0.88\linewidth]{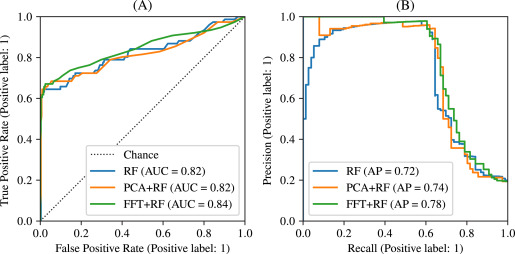}
    \caption{The ROC curve (left) and precision-recall curve (right) of the RF classifier baseline, RF with PCA and RF with FFT models applied on the beam loss dataset. (From~\cite{revsvcivc2022improvements}, Fig. 7)}
    \label{fig:sns}
\end{figure*}

In a previous study, \citet{rescic2020predicting} have gone through a holistic research of ML classification methods, including logistic regression, $k$ nearest neighbours, tree-based methods, support vector machines and MLPs, and achieves almost 92\% accuracy in identifying bad pulses from a MLP model combined with parameter tuning and data refining. However, only the pulses right before and right after the trips are taken and labeled, and nearly 8\% of daily good pulses are incorrectly predicted as bad. Their following-up work~\cite{revsvcivc2022improvements} improves the result with a more complete dataset, and introduces fast Fourier transform (FFT) for feature extraction and principal component analysis (PCA) for dimensionality reduction. \SI{26}{} pulses before the trips --- instead of only one previously --- are taken as bad pulses (labeled as \emph{Before} pulses) that lead to failure, \SI{2}{} pulses after the trips (labeled as \emph{After} pulses) are taken as good pulses, and pulses in normal operation without trips are also taken and labeled as \emph{Notrip} pulses. Both the \emph{After} pulses and the \emph{Notrip} pulses are classified respectively against the \emph{Before} ones. By analysing their distance to next trips and comparing the classification results, the newly taken \emph{Notrip} pulses are considered to be more representative for normal operation. Therefore the authors decide to focus on the \emph{Before}--\emph{Notrip} classifier. The best-performing Random Forest model together with PCA achieves 96\% accuracy and 61\% recall. By further leveraging classification-threshold and improvement techniques, the classifier could in principle reach a strict 0\% false-positive rate, at the cost of a true-positive rate of less than 58\%. Figure~\ref{fig:sns} shows the receiver operating characteristics (ROC) curve and the precision-recall curve of three models on the beam loss dataset, which is taken around machine trips where beam loss occurred. Both types of curves are generated by leveraging the classification threshold from 0 to 1. The ROC curve shows the true-positive rate against the false-positive rate; the uppermost left curve is optimal and has the greatest area under the curve (AUC) value. The average precision (AP) is calculated over the full threshold range.

All classifiers successfully predict the accelerator failures inside \SI{4}{\milli\second}, far less than the available time budget of \SI{16}{\milli\second}. This enables SNS to implement the model in real-time operation, and invoke mitigation techniques in field-programmable gate arrays (FPGAs) to realize the inhibition of pulses and resetting of the machine.

\begin{figure*}
    \centering
    \subfloat{\includegraphics[width=0.45\linewidth]{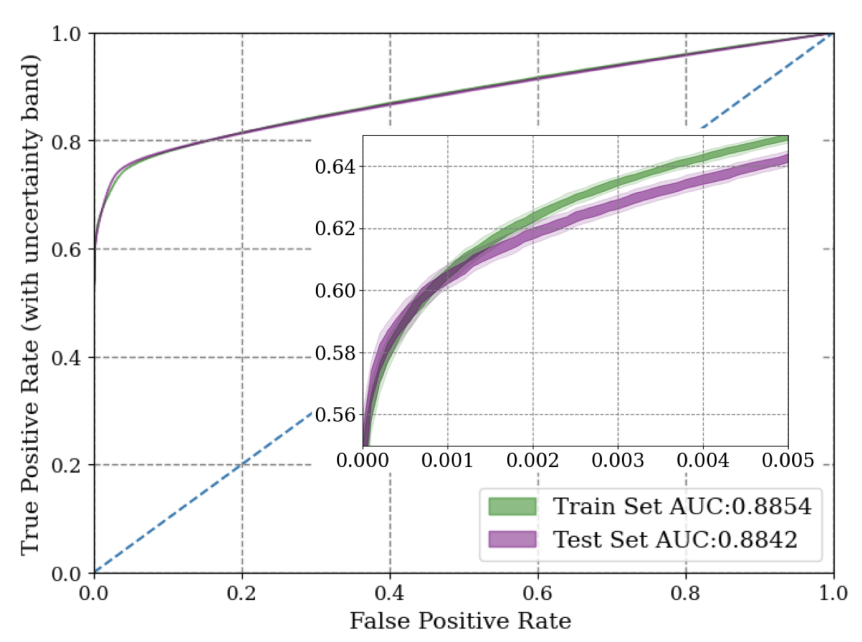}}
    \subfloat{\includegraphics[width=0.45\linewidth]{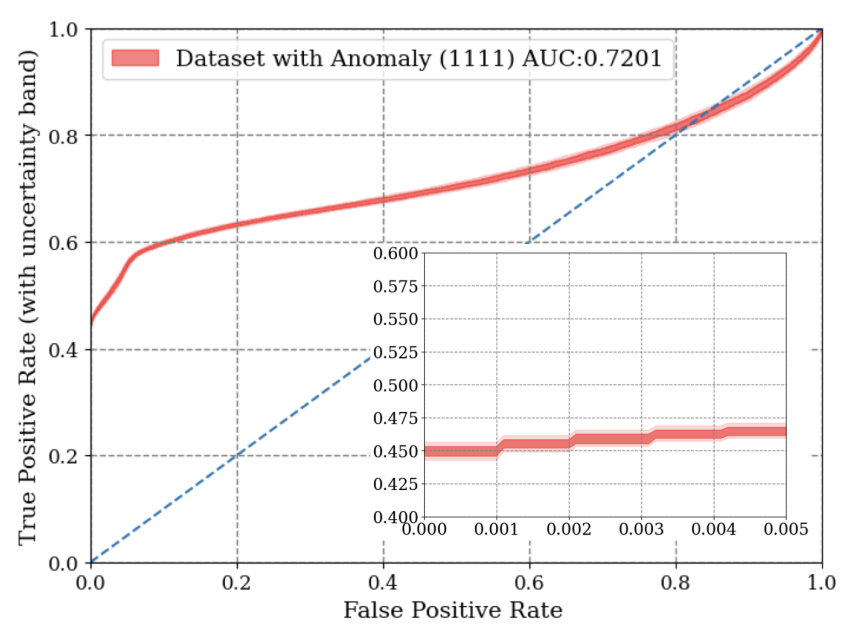}}
    
    \caption{The ROC curves for training and testing with the same (left) and different (right) errant types. (From~\cite{blokland2021uncertainty}, Figs 14 \& 15)}
    \label{fig:siamese}
\end{figure*}

% siamese
Another inspiring study using the same DCM data from SNS realizes uncertainty-aware anomaly detection of the pulses. \citet{blokland2021uncertainty} build a Siamese neural network~\cite{koch2015siamese} to distinguish two types of errant pulses from the normal pulses by ranking their similarity. While keeping the false-positive rate below the established 0.05\% limit, the true-positive rate increases to more than 64\% when training and testing with the same errant type, and also reaches 45\% in cross-type testing, as shown in Figure~\ref{fig:siamese}.

Starting from the current offline model validation result, the authors have been working towards online prediction together with real-time implementation, where they compare the incoming pulse with a series of past pulses using the Siamese network, make decisions based on their similarity level, and abort the predicted errant beam to reduce system downtime.

\subsection{Interlock forecasting of the High Intensity Proton Accelerators}\label{sec:hipa}

HIPA produces a proton beam of nearly \SI{1.4}{\mega\watt} power, which makes it one of the most powerful proton cyclotron facilities in the world~\cite{reggiani2020improving}. The interlock system is part of the Machine Protection System that immediately shuts off the beam whenever some monitor signal exceeds the safety limit. However, such shutdowns may lead to abrupt operational changes and a substantial loss of beam time. \citet{li2021novel} propose to build a forecasting model of the interlocks. Once the model reports an incoming interlock, the suggested recovery operation to reduce the beam current by 10\% would be applied, which could potentially circumvent the interlocks from happening, thus save beam time for the users.

\begin{figure*}
    \centering
    \includegraphics[width=0.95\linewidth]{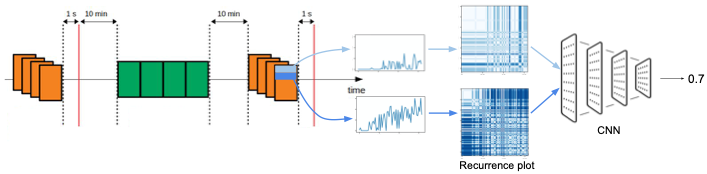}
    \caption{The RPCNN model structure. The positive and negative classes of samples are taken either close to (orange) or far from (green) the interlocks. Each of the \SI{376}{} time series are transformed into recurrence plots and fed into the CNN. The model output is a probability value. (Adapted from~\cite{li2021novel}, Figs 5 \& 8)}
    \label{fig:rpcnn}
\end{figure*}

The dataset is composed of \SI{376}{} process variables from the Experimental Physics and Industrial Control System (EPICS), which are recorded at a \SI{5}{\hertz} frequency. The problem is formulated as binary classification of two classes of samples. The positive class consists of \emph{interlock samples} that are taken at \SI{1}{\second} to \SI{12}{\second} before the interlocks, shown in orange colour in Figure~\ref{fig:rpcnn}. This is how the concept of forecasting is embedded here, just like taking the pulses before the trips in the SNS case. The negative class consists of \emph{stable samples} that are taken in the middle between two adjacent interlocks with a buffer region of \SI{10}{\minute} on both sides, to represent stable operating states, shown in green in Figure~\ref{fig:rpcnn}. The window length is a trainable parameter to be decided in the model. 

The authors develop a recurrence plot-convolution neural network (RPCNN) model for the classification task. Each one-dimensional time series of the input is transformed into a two-dimensional \emph{recurrence plot} to extract finer dynamical patterns. Then the plots are trained with a \emph{convolutional neural network} (CNN), which is an established and powerful method in image processing. In practice, the recurrence plots are produced internally by a custom \emph{recurrence plot layer} before the convolutional and max-pooling layers. This procedure prevents the recurrence plots from being generated and stored explicitly beforehand, and also allows the optimization of the plots on the fly. The output is a score $y \in [0,1]$, indicating the probability that the incoming sample belongs to the positive class, thereby forecasting an interlock.

\begin{figure}
    \centering
    \includegraphics[width=0.78\linewidth]{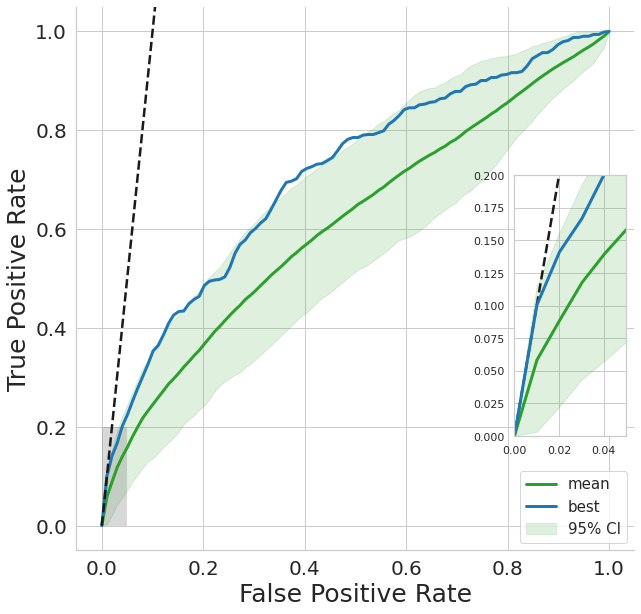}
    \caption{The ROC curves of an ensemble of RPCNN classifiers. The blue line shows the best classifier with AUC=0.71, the green line is the mean curve, and the shaded area is the $95\%$ confidence interval of different model initialization profiles. The dashed line is the separatrix, left of which there is a positive \emph{beam time saved}. The inset is a zoom-in on the grey shaded region. (From~\cite{li2021novel}, Fig. 11)}
    \label{fig:rpcnn_roc}
\end{figure}

The authors choose the best-performing model based on a custom metric they call \emph{beam time saved}, which computes potential time saved by invoking the recovery operation back on the machine. As the recovery operation would cost an equivalent of \SI{6}{\second} of beam-time loss per interlock, false positives need to be strictly controlled in order to reach a bonus in \emph{beam time saved}. Therefore, the resulting classifier has a true-positive rate of 4.9\%, together with an extremely low false-positive rate of 0.17\%, and it can potentially save \SI{0.5}{\second} of beam time per interlock. Figure~\ref{fig:rpcnn_roc} shows the best and mean ROC curves of RPCNN classifiers with random initialization, as well as their uncertainties.

To alleviate the limitations of false positives, the authors further study the input channels and discover from statistical tests that a significant difference is only present inside \SI{0.4}{\second} before the interlocks. A preliminary study with a linear least absolute shrinkage and selection operator (LASSO) model in which only single timestamps are used and the positive samples are pushed closer to the interlocks, showed improvements in both classification power and stability. The \emph{beam time saved} metric is also modified to fit the continuous real-time context, and the new model is shown to potentially save around \SI{6}{\minute} beam time in a day.
\subsection{Power supply trips prediction in the Advanced Photon Source storage ring}
The Advanced Photon Source (APS) at Argonne National lab provides ultra-bright X-ray beams of \SI{7}{\giga\eV} for advanced research. Trips in the magnet power supplies of the storage ring are highly undesirable, as they would cause complete electron beam loss and interrupt user experiments instantly. Because the trips are rare events with diverse triggering mechanisms, over-fitting would become inevitable in supervised learning, and labelling them as one class cannot reflect reality either. Therefore, \citet{lobach2022machine} focus on unsupervised anomaly detection methods that train on normal operation data and identify trip precursors by measuring their level of deviation.

For the temperature anomalies caused by valve faults in the water-cooling system, the authors apply the \emph{spectral residual saliency detection} method on the temperatures of \SI{680}{} power supplies in a time window of \SI{3}{\hour}. The anomalies clearly stand out on the saliency maps, and the model successfully gives warnings up to \SI{30}{\minute} in advance. 

The authors achieve an even earlier advanced warning of \SI{1}{\hour} by training an \emph{autoencoder} on normal operation data of the temperatures from \SI{40}{} averaged power supplies and tracking the reconstruction error at each time step. If only one power supply temperature is considered as input, an autoencoder trained and tested on sliding windows of \SI{20}{\minute} would even give the warning \SI{6}{\hour} before the trip happens.

Following the above success, the authors employ the autoencoder approach again on power supply current anomalies and obtain a preliminary result of 20\% true-positive rate. While there is still much room for improvement in the current approach, they already show great potential in early-enough warning and possibilities for preventive action.
\subsection{Real-time cavity fault prediction
at the Continuous Electron Beam Accelerator Facility}

\begin{figure}
\centering
\includegraphics[width=\linewidth]{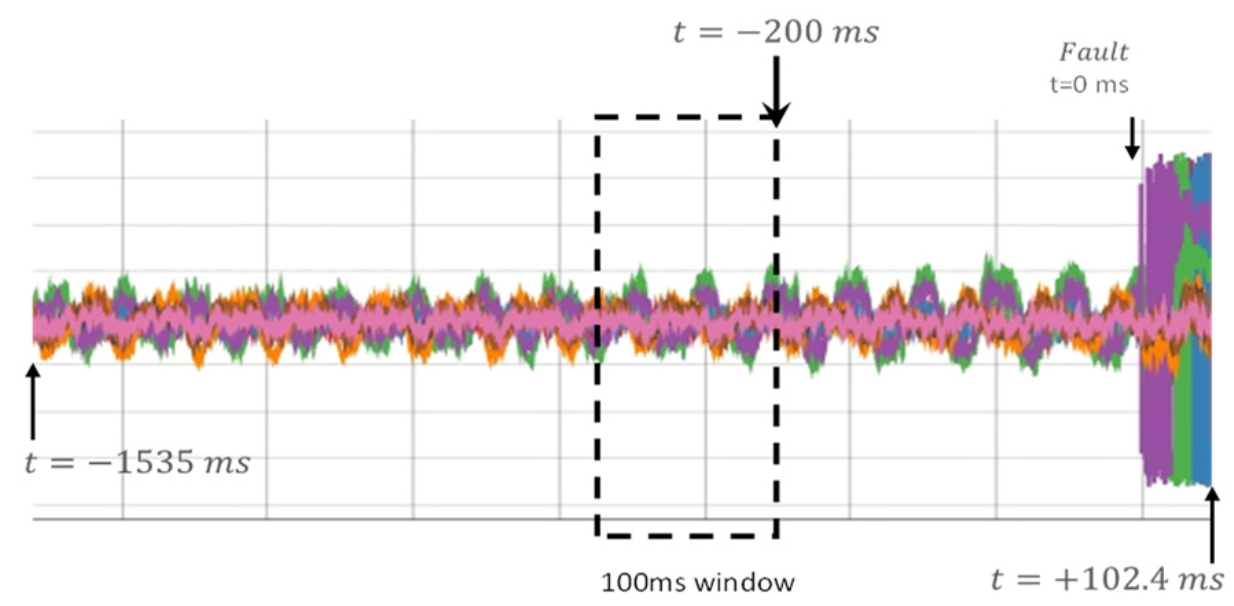}
\caption{Recorded waveforms and window taking of the CEBAF SRF fault forecasting model. (From~\cite{rahman2022real}, Fig. 2)}
\label{fig:cebaf_window}
\end{figure}

The Continuous Electron Beam Accelerator Facility
(CEBAF) at Jefferson Lab is a high power, continuous
wave recirculating Linac whose peak energy reaches \SI{12}{\giga\eV}. The cryomodules that provide the energy gain are composed of superconducting radio-frequency (SRF) cavities, and the machine experiences frequent downtimes caused by various types of SRF faults --- on average \SI{4.1}{} times in an hour, which amounts to about \SI{1}{\hour} beam time loss per day. \citet{tennant2020superconducting} have proposed successful machine learning models that realize fast classification of SRF faults. In a recent follow-up work~\cite{rahman2022real}, the authors go on to build deep learning based forecasting models for such faults. The data acquired from CEBAF contains waveforms of \SI{17}{} RF signals, \SI{4}{} of which are used in the study. The sample interval is \SI{0.2}{\milli\second}, and each waveform lasts for \SI{1637.4}{\milli\second} including \SI{1535}{\milli\second} before and \SI{102.4}{\milli\second} after the fault onset. The forecasting of impending faults is also formulated as binary classification between two classes of \SI{100}{\milli\second} windows -- the \emph{stable class} taking from normal running conditions, and the \emph{pre-fault class} taking at \emph{lead time} $h \in [200, 100, 50, 20, 10, 5, 0] \,\SI{}{\milli\second}$ before the fault onset. Figure~\ref{fig:cebaf_window} shows an example of the recorded waveforms of the \SI{4}{} signals and illustrates the $h=\SI{200}{\milli\second}$ window taking of the \emph{pre-fault} class. 

\begin{figure}
    \centering
    \includegraphics[width=0.9\linewidth]{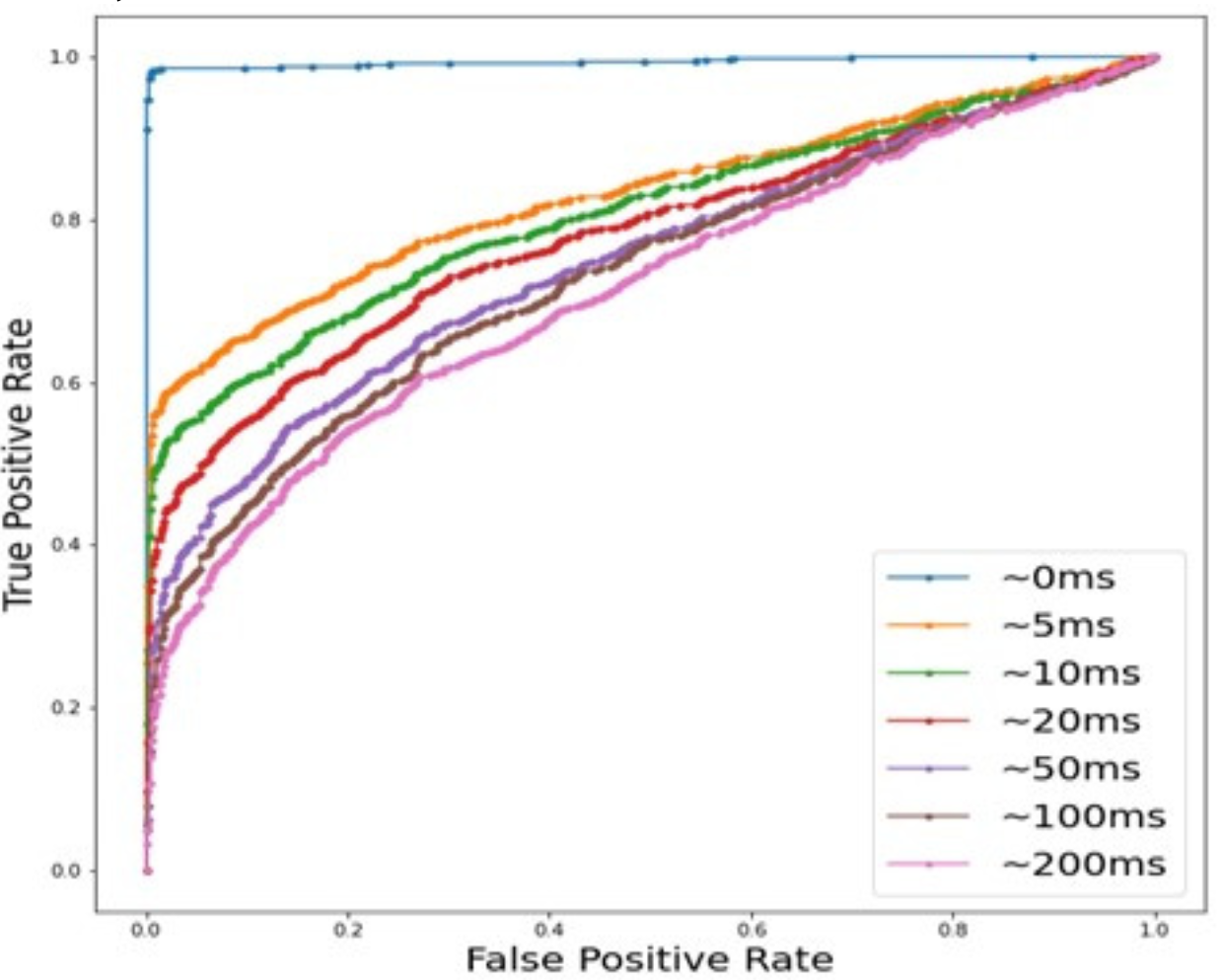}
    \caption{ROC curves of classifying samples taken at different times before the SRF faults against normal samples. (From~\cite{rahman2022real}, Fig. 5)}
    \label{fig:cebaf_roc}
\end{figure}

Adopting the U-Net architecture~\cite{ronneberger2015u}, the binary classification model is trained only with \emph{normal class} samples, aims to output similar normal samples by minimizing the reconstruction loss. During testing, a \emph{pre-fault} sample would lead to larger reconstruction error thus indicates its abnormality. Figure~\ref{fig:cebaf_roc} lists the ROC curves for different time before the fault onset. The closest prediction at $h=\SI{5}{\milli\second}$ reaches AUC $=0.83$, whereas the earliest prediction at $h=\SI{200}{\milli\second}$ has AUC $=0.71$. Though it is evident that the performance declines with longer prediction time, the results have specified the timescales --- about a few hundred milliseconds --- for possible mitigating operations.

Furthermore, the authors build a subsequent multi-class classification network for fault-type identification from the outputs of the previous model that are classified as upcoming faults. Results show that fast-developing faults are harder to identify than slow-developing ones if samples are taken far from the faults. This echoes with the discovery of HIPA interlocks mentioned in \ref{sec:hipa} that the forecasting of abrupt anomalies is challenging.

\subsection{Beam loss prediction at the Large Hadron Collider}\label{sec:cern}
Beam losses in the Large Hadron Collider (LHC) at CERN are mostly occurring in the collimation system to remove particles with excessively high oscillation or momentum. The loss level is manually optimized by multiple control variables, including vertical and horizontal tunes, and currents in the focusing and defocusing magnets along the collider. It is therefore crucial to model the beam loss from those control variables for better machine operation. However, there is the problem of generalization shown by a previous study~\cite{coyle2018machine}, when the model is trained on previous LHC fills and then applied to fills of another year. \citet{krymova2022accel} propose an \emph{autoregressive} approach that factors in past value of losses to alleviate the problem. To take advantage of the efficient inference procedure of \emph{state-space models}, the authors transcribe the autoregressive formulation into a Kalman filter formulation, and established several model variants with different inputs and outputs. In addition to the control variables, measurements of emittance and heatload sum are also taken into account as possible inputs. By making the parameter matrices dependent on the control variables, the authors manage to build a nonlinear model including cross terms between different inputs. For the estimation of model parameters, a customized expectation-maximization algorithm is implemented.

\begin{figure}
\centering
\subfloat{\includegraphics[width=1\linewidth]{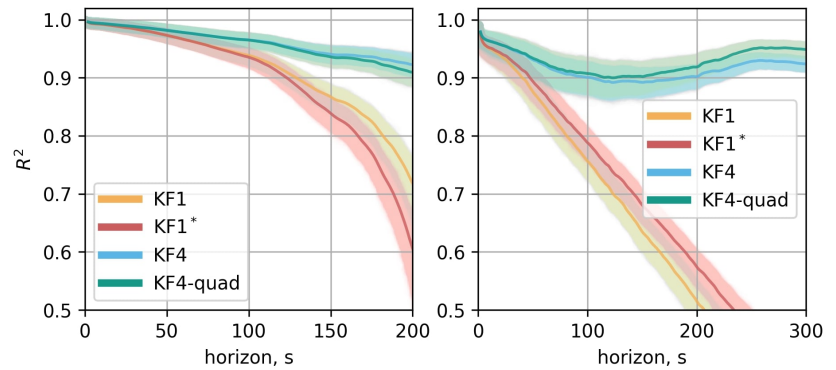}}

\subfloat{\includegraphics[width=0.5\linewidth]{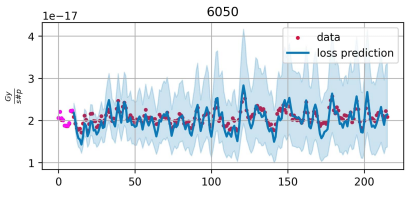}}
\subfloat{\includegraphics[width=0.5\linewidth]{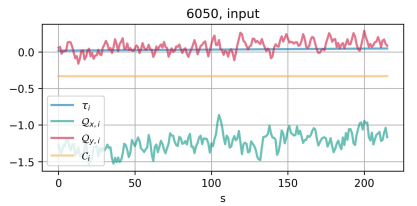}}
\caption{Mean and envelope of 1000 bootstrapped estimates of $R^2$-score against forecast horizons for the 2017 training set (upper left) and the 2018 testing set (upper right), the prediction result of Fill 6050 from the KF4 model with $2\sigma$ confidence band (lower left), and the corresponding inputs (lower right). The horizons are \SI{200}{\second} in 2017 and \SI{300}{\second} in 2018. (From~\cite{krymova2022accel}, Figs. 8 \& 10)}
\label{fig:cern}
\end{figure}

Figure~\ref{fig:cern} shows the $R^2$-score of predictions against forecast horizons from four models, together with the prediction result and corresponding input series of an example fill from the model $KF4$, which involves non-controlled measurements and additional output components. The model parameters are estimated from a training set with data taken in 2017 and evaluated on a 2018 testing set. The excellent performance on long horizons have provided encouraging evidence that a carefully designed model can capture the global trend and simultaneously establish a relation with inputs.
 
\subsection{Proposed energy stability prediction at the High Repetition rate Electron Scattering beamline}

The High Repetition rate Electron Scattering (HiRES) beamline at Lawrence Berkeley National Laboratory is a state-of-the-art compact machine for \SI{}{\mega\hertz} ultrafast electron diffraction (UED) pulses. \citet{scheinker2013model,scheinker2018constrained} has previously proposed extremum seeking (ES) as an optimization technique that realizes automatic and model-independent tuning for accelerator parameters. It is also proved to be robust against drift that brings system outside the training range~\cite{scheinker2021adaptive}.

With the help of the high-resolution FPGA-enabled feedback system, HiRES is shown to be stable against jitters and it is established that the machine could reach an energy stability of $\Delta E/E=5\times 10^{-5}$ on short timescales. However, unknown parameter drifts on longer timescales could magnify the energy deviation by more than 10 times. A non-static diagnostic model that involves time evolution is therefore proposed, which would be integrated with the previous optimization technique to establish a novel suite of  ML-based adaptive control systems for intelligent feedback.

\begin{figure}
    \centering
    \includegraphics[width=\linewidth]{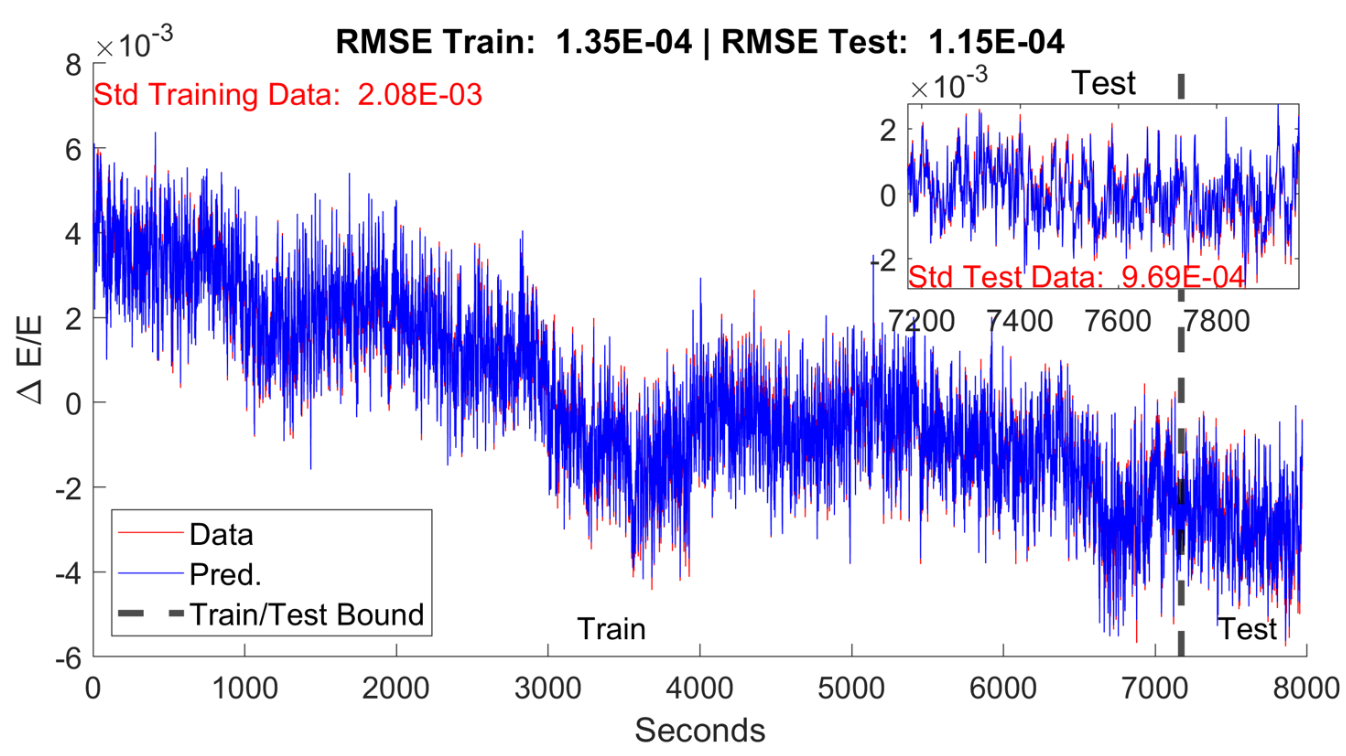}
    \caption{Linear regression result of energy stability. The dashed line splits the training (left) and test (right) data. (From~\cite{cropp2022toward}, Fig. 6)}
    \label{fig:hires_result}
\end{figure}

\begin{figure}
\centering
\subfloat[Not use target values to predict.]{\includegraphics[width=1\linewidth]{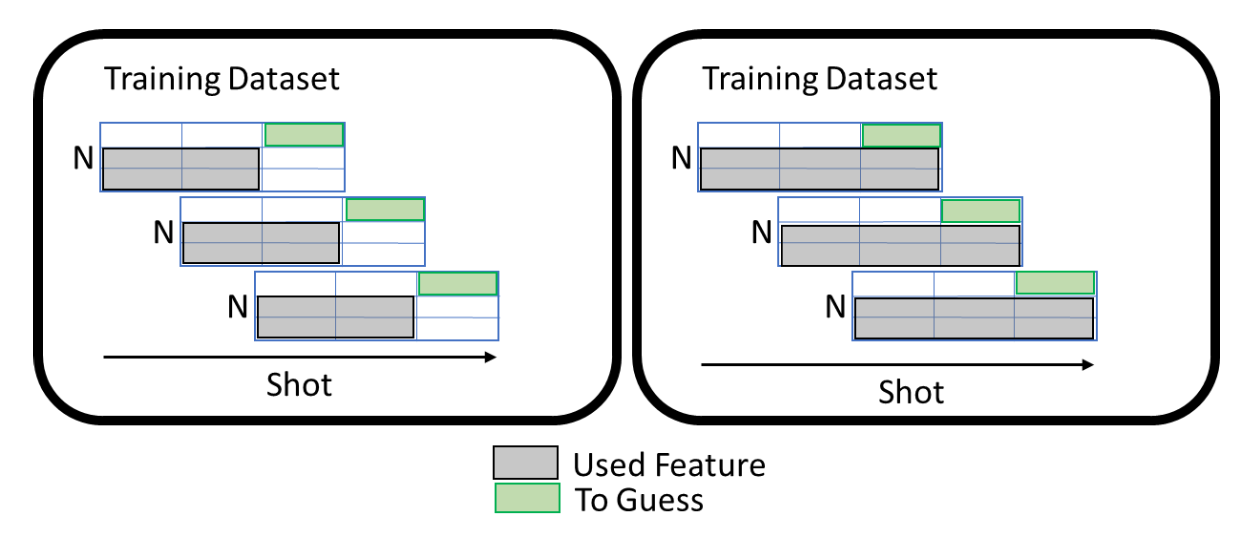}}

\subfloat[Use both predictors and target to predict.]{\includegraphics[width=1\linewidth]{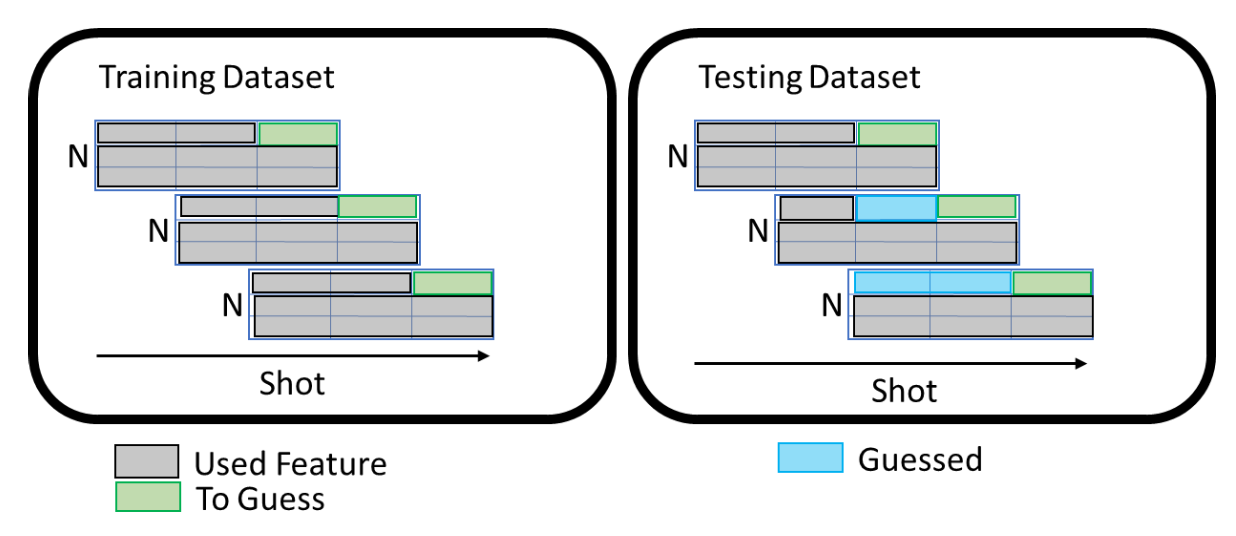}}

\caption{Possible set-up for time series prediction. (a) Predictors (grey) and target (green) are different. The left and right plots show whether to include the present predictors to predict the present target. (b) Target is included as predictor. In the right plot, past predictions of the target (blue) are used in the testing dataset. (From~\cite{cropp2022toward}, Figs 4 \& 5)}
\label{fig:time_window}
\end{figure}

To infer the energy stability $\Delta E/E$, \citet{cropp2022toward} have set out to predict the beam $x$-position from the amplitude and phase of cavity probes, RF power, and laser properties on a virtual cathode camera. Currently results such as the one shown in Figure~\ref{fig:hires_result} are still achieved by simple linear regression at each time step, without considering past information. According to the authors, this preliminary regression result has already helped to ensure much less fluctuation than in the original hardware feedback system. Starting from this, the authors consider time series forecasting techniques and conceive various possible ways to incorporate time evolution into the model, as shown in Figure~\ref{fig:time_window}. Following the result presented in Section~\ref{sec:cern}, such a model involving time would be a natural oppressor for long-term drifts once properly defined. %Further incorporation with time series forecasting is planned.

\section{Discussion and Conclusion}\label{sec:disc}
\subsection{Data related challenges}
The main problem reported in ML applications for particle accelerator scenarios is the data --- although a large quantity of data is available, there is still a long process to go through until an instructive or even deployable model can be being built. First, decisions need to be made about which data source is to be taken, how to extract and record the data, and about the merging of possibly different logging systems. Then the storage format needs to be decided and unified, which has to meet also the requirement for extensibility and easy query. Once the model is built, further issues arise such as accommodating the real-time input, online learning, model updating, storage of temporary results, and finally, feeding the operations invoked by the model output back to the machine. Such issues have already raised considerable interest and effort in the community. For instance, \citet{kafkes2021boostr} have published the Booster Operation Optimization Sequential Time-series for Regression (BOOSTR) dataset, which is composed of \SI{15}{\hertz} cycle-by-cycle multivariate time series of readings and settings from devices of the Rapid-Cycling Synchrotron at Fermilab. Such an attempt is very encouraging, as it addresses the problem from the root, while also aiming to create a more open and inclusive system to promote the data-driven research in the particle accelerator community.

\subsection{Insight from remaining useful life prediction}
The remaining useful life (RUL) prediction has been an important research topic in the predictive maintenance field, aiming to detect possible defects early and thus to identify and apply the required maintenance activities such that possible breakdowns are avoided. Instead of predicting an anomaly score for the input signals at every time step, the output for RUL predictions is the duration from the current time until the nearest failure. According to~\citet{kang2021remaining}, ML techniques have also spawned many new attempts in this area, and the problem settings are also migratable to particle accelerator control, despite the timescale difference. However, RUL prediction is mainly applied in device degradation, which possesses a clear gradual change curve. There are even physical models established to explain such effects. Equivalent models lack in accelerator control, and preliminary trials have shown that such methods do not yield satisfactory performance on abrupt failures.

%conclusion
In conclusion, there are bright prospects for the application of data-driven time series forecasting techniques in problems related to particle accelerators, especially in control and diagnostics. The field would benefit from an extension of current research, an increasing attention to data quality, innovative insights from similar fields and more intense exchange. In this way time series forecasting models will emerge that are more tightly tailored to particle accelerator scenarios.

%Here is an example usage of the two main commands (\verb+citet+ and \verb+citep+): Some people thought a thing \citep{li2010applications, makridakis2018statistical} but other people thought something else \citep{li2010applications}. Many people have speculated that if we knew exactly why \citet{chatfield2000time} thought this\dots

%\bibliographystyle{unsrt}
%\bibliographystyle{plainnat} % author-year citation
\bibliographystyle{apsrev4-2}
\bibliography{references}
\end{document}